\documentstyle[12pt,preprint,aps]{revtex}

\begin{document}

\title{Rotating Hele-Shaw cells with ferrofluids}

\author{Jos\'e A. Miranda\footnote{e-mail:jme@lftc.ufpe.br}}
\address{Laborat\'{o}rio de F\'{\i}sica Te\'{o}rica e Computacional,
Departamento de F\'{\i}sica,\\ Universidade Federal de Pernambuco, 
Recife, PE  50670-901 Brazil}
\date{\today}
\maketitle

\begin{abstract} 

We investigate the flow of two immiscible, viscous fluids 
in a rotating Hele-Shaw cell, when one of the fluids 
is a ferrofluid and an external magnetic field is applied. 
The interplay between centrifugal and magnetic forces in 
determining the instability of the fluid-fluid 
interface is analyzed. The linear stability analysis of the problem 
shows that a non-uniform, azimuthal magnetic field, applied 
tangential to the cell, tends to stabilize the interface. We 
verify that maximum growth rate selection of initial patterns is 
influenced by the applied field, which tends to decrease 
the number of interface ripples. We contrast these results with the 
situation in which a uniform magnetic field is applied normally 
to the plane defined by the rotating Hele-Shaw cell.

\end{abstract}


When a fluid is pushed by a less viscous one in a narrow 
space between two parallel plates (the so-called Hele-Shaw cell), 
the well-known Saffman-Taylor instability phenomenon arises~\cite{Saf}, 
which takes the form of fingering~\cite{Rev}. Traditionally, experiments 
and theory focus on two basic Hele-Shaw flow geometries: 
(i) rectangular~\cite{Saf} and radial~\cite{Pat}. In rectangular 
geometry cells the less viscous fluid is pumped against the more viscous 
one along the direction of the flow. Meanwhile, in the radial geometry 
case, the less viscous fluid is injected to invade the more viscous one, 
through an inlet located on the top glass plate. In both geometries, the 
viscosity-driven instability leads to the formation of beautiful 
fingering patterns.

In recent years, the quest for new morphologies and 
richer dynamic behavior resulted in a number of modifications of the 
classic Saffman-Taylor setup~\cite{McC}. An interesting variation 
of the traditional viscosity-driven fingering instability 
is the investigation of radial Hele-Shaw flows 
in the presence of centrifugal driving. The inclusion of centrifugal 
forces can be considered by rotating the cell, with constant 
angular velocity, around an axis perpendicular to the plane 
of the flow. In this case, the interface 
instability can be driven by the density difference between the fluids. 
In the late 1980s Schwartz~\cite{Sch} performed the linear stability 
analysis of the rotating cell problem, in the limits of high 
density and viscosity contrast. More recently, Carrillo et 
al.~\cite{Car} studied, both theoretically and experimentally, 
flow in a rotating Hele-Shaw cell in arbitrary density and viscosity 
contrast. They extended the linear analysis performed in~\cite{Sch} by 
considering that the inner fluid is injected in the cell through a 
hole at the center of rotation, with constant injection rate. 
The linear growth rate calculated in~\cite{Car} shows that 
the interface instability can be driven by both the density difference 
and the viscosity contrast between the fluids.  Their experimental 
results supported their theoretical analysis. Carrillo and 
co-workers also examined the radial displacement of a rotating 
fluid annulus, bound by a second fluid, in 
stable~\cite{Car2} and unstable~\cite{Car3} regimes. In another 
interesting work, Magdaleno et al.~\cite{Magda} applied a 
conformal mapping technique to argue that the effects of rotation 
can be used to prevent cusp singularities in zero surface 
tension Hele-Shaw flows.

Another stimulating modification of the traditional Saffman-Taylor 
problem in non-rotating Hele-Shaw cells considers the interface 
morphology when one of the fluids is a ferrofluid~\cite{Ros}, and 
an external magnetic field is applied perpendicular to the cell 
plates. Ferrofluids, which are colloidal suspensions of 
microscopic permanent magnets, respond paramagnetically to 
applied fields. As a result of the ferrofluid interaction with 
the external field, the usual viscous fingering instability is 
supplemented by a magnetic fluid instability~\cite{Ros}, resulting 
in a variety of new interfacial behaviors~\cite{Tse,Lan5,Jac5,Eu,F1,C1}.

The richness of new behaviors introduced by both 
rotation and magnetic field into the traditional Saffman-Taylor 
problem, motivated us to analyze the situation in which these 
two effects are simultaneously present. In this work we perform 
the linear stability analysis for flow in a rotating Hele-Shaw cell, 
assuming that one of the fluids is a ferrofluid and that 
a magnetic field is applied. First, we consider the new 
situation in which a {\it non-uniform}, 
azimuthal, in-plane field is applied. The competition between 
rotation and magnetic field is analyzed. We show the 
azimuthal magnetic field provides a new mechanism for 
stabilizing the interface. It is the first time such a field 
induced, stabilizing mechanism is proposed, and proved to be effective 
for flow in rotating Hele-Shaw cells. Through the analysis of the 
maximum growth rate, we verify that the azimuthal magnetic field acts to decrease the number of interface ripples. Finally, we contrast these 
results with the destabilizing situation in which a {\it uniform} 
magnetic field is applied normally to the plane defined by 
the rotating Hele-Shaw cell.

Consider a Hele-Shaw cell of thickness $b$ containing two immiscible, 
incompressible, viscous fluids (see figure 1). Denote the densities and 
viscosities of the inner and outer fluids, respectively as $\rho_{1}$,
$\eta_{1}$ and $\rho_{2}$, $\eta_{2}$. The flows in fluids $1$ and $2$ 
are assumed to be irrotational. Between the two fluids there 
exists a surface tension $\sigma$. We assume that the inner fluid 
is the ferrofluid (magnetization $\vec M$), while the outer 
fluid is non-magnetic. During the flow, the fluid-fluid interface 
has a perturbed shape described as ${\cal R}= R + \zeta(\theta,t)$, 
where $\theta$ represents the polar angle, and $R$ is the 
radius of the initially unperturbed interface. 

In order to include centrifugal forces, we allow the cell 
to rotate, with constant angular velocity $\Omega$, about 
an axis perpendicular to the plane of the flow (figure 1). To include
magnetic forces, we consider the action of an external magnetic 
field $\vec H$, produced by a long, straight wire 
carrying a current $I$, directed along the axis of 
rotation. By Ampere's law, it can be shown that the 
steady current $I$ produces an azimuthal magnetic field external 
to the wire $\vec H =(I/2 \pi r) ~\hat{\theta}$, 
where $r$ is the distance from the wire, and $\hat{\theta}$ is 
the unit vector pointing in the direction of increase of $\theta$. 

Following the standard approximations 
used by Rosensweig~\cite{Ros} and others~\cite{Tse,Lan5,Jac5} 
we assume that the ferrofluid magnetization 
$\vec M$ is collinear with the external field $\vec H$ and that the 
influence of the demagnetizing field is neglected. It is also 
assumed that the ferrofluid is electrically nonconducting and that the 
displacement current is negligible. For the quasi two-dimensional 
geometry of a Hele-Shaw cell, the three
dimensional flow may be replaced with an equivalent two-dimensional
flow ${\vec v}(x,y)$ by averaging over the $z$ direction perpendicular
to the plane of the Hele-Shaw cell. Imposing no-slip boundary
conditions and a parabolic velocity profile one derives Darcy's law
for ferrofluids in a Hele-Shaw cell~\cite{Jac5,Tse2}, which  must be 
augmented by including centrifugal forces,
\begin{equation}
\label{Darcy}
\eta \vec v= -\frac{b^{2}}{12} \left \{ \vec \nabla p -
\frac{1}{b} \int_{-b/2}^{+b/2} \mu_{0}(\vec M \cdot \vec \nabla) \vec H dz
- \rho \Omega^{2} r ~\hat{r} \right \},
\end{equation}
where $p$ is the hydrodynamic pressure, $\mu_{0}$ is the free-space 
permeability, and $\hat{r}$ denotes 
a unit vector pointing radially outward. Equation~(\ref{Darcy}) 
describes non-magnetic fluids by simply dropping
the terms involving magnetization.

Since the velocity field ${\vec v}$ is irrotational, it is convenient
to rewrite equation~(\ref{Darcy}) in terms of velocity potentials. We
write $\vec v=-\vec\nabla\phi$, where $\phi$ defines the velocity
potential. Similarly, we rewrite the magnetic body force in equation~(\ref{Darcy}) as $\mu_{0}(\vec M \cdot \vec \nabla) \vec H= 
\mu_{0}M \vec \nabla H=\vec \nabla \Psi$, where we have 
introduced the scalar potential
\begin{equation}
\label{Psi2}
\Psi=\mu_{0} \int M(H) dH= \frac{\mu_{0} \chi H^{2}}{2}, 
\end{equation}
with $M=M(H)=\chi H$, $\chi$ being a constant magnetic susceptibility.

With the definitions of $\phi$ and $\Psi$ we notice that both 
sides of equation~(\ref{Darcy}) are recognized as gradients of 
scalar fields. After integrating both sides of equation~(\ref{Darcy}), 
we evaluate it for each of the fluids on the interface. Then, we 
subtract the resulting equations from each 
other, and divide by the sum of the two fluids' viscosities to get 
the equation of motion
\begin{equation}
\label{difference}
A \left ( {{\phi_2 + \phi_1}\over{2}} \right ) + \left ( \frac{\phi_2 - \phi_1}{2} \right ) = -{{b^{2}}\over{12(\eta_{1} + \eta_{2})}} \Bigg \{ \sigma \kappa - \Psi +  \frac{1}{2} (\rho_{2} - \rho_{1}) \Omega^{2} r^{2} \Bigg \}.
\end{equation}
To obtain~(\ref{difference}) we have used the pressure boundary
condition $p_{1} - p_{2}=\sigma\kappa$ at the interface, where
$\kappa= \left \{[ r^2 + 2 (\partial r / \partial \theta)^2 - r (\partial^2 r / \partial \theta^{2})] / [ r^2 + (\partial r / \partial \theta)^2]^{3/2} \right \}$ denotes the interfacial curvature in the plane
of the Hele-Shaw cell. The dimensionless parameter $A=(\eta_{2} -
\eta_{1})/(\eta_{2} + \eta_{1})$ is the viscosity contrast.

For the purpose of the following linear analysis, we perturb the 
interface with a single Fourier mode
\begin{equation}
\label{expansion}
\zeta(\theta,t)=\zeta_{n}(t) 
\exp{(i n \theta)}, \,\,\,\ n=0,1,2, $...$
\end{equation}
The velocity potential for fluid $j$ ($j=1,2$ indexes the inner and 
outer fluids, respectively), $\phi_{j}$, obeys Laplace's equation
$\nabla^{2}\phi_{j}=0$ and can be written as
\begin{equation}
\label{phi1-2}
\phi_{j}= \phi_{j}^{0} + \phi_{j n} \left ( \frac{R^{n}}{r^{n}} \right )^{(-1)^{j}} \exp(i n \theta),
\end{equation}
where $\phi_{j}^{0}$ are independent of $r$ and $\theta$.

We need additional relations expressing the velocity potentials 
$\phi_{j}$ in terms of the perturbation amplitudes $\zeta$, 
in order to conclude our derivation and close 
equation~(\ref{difference}). To find these, we considered the kinematic
boundary condition~\cite{Ros}, which refers to the continuity of the normal 
velocity across the interface. Inserting expression~(\ref{expansion}) 
for $\zeta(\theta,t)$ and~(\ref{phi1-2}) for $\phi_j$ into the 
kinematic boundary condition, we solved for $\phi_{jn}$ consistently 
to first order in $\zeta$ to find
\begin{equation}
\label{phi1t}
\phi_{jn} = (-1)^{j} \frac{R}{n}\dot{\zeta}_{n},
\end{equation}
where the overdot denotes total time derivative.

Substitute expression~(\ref{phi1t}) for $\phi_{jn}$ into equation of motion~(\ref{difference}),
and again keep only linear terms in the perturbation amplitude. 
This procedure eliminates the velocity potentials from
equation~(\ref{difference}), and we obtain the differential  
equation for the perturbation amplitudes 
$\dot{\zeta}_{n}=\lambda(n) \zeta_{n}$, implying that the relaxation 
or growth of the mode $n$ is proportional to the factor 
$\exp[\lambda(n) t]$, where
\begin{equation}
\label{dispersion}
\lambda(n)=\frac{b^{2} \sigma n}{12(\eta_{1} + \eta_{2})R^{3} } \left [ N_{\Omega} - N_{B} - (n^{2} - 1) \right ]
\end{equation}
is the linear growth rate. We define the dimensionless parameters $N_{\Omega}=[R^{3} (\rho_{1} - \rho_{2}) \Omega^{2}]/\sigma$, and 
$N_{B}=\mu_{0} \chi I^{2}/(4 \pi^{2} \sigma R)$ as the rotational 
and magnetic bond numbers, respectively. $N_{\Omega}$ ($N_{B}$) measures the relative strength of centrifugal (magnetic) and capillary effects.

Inspecting equation~(\ref{dispersion}) for the linear growth rate 
$\lambda(n)$ we observe the interplay of rotation, 
magnetic field and surface tension in 
determining the interface instability. If $\lambda(n) > 0$ the 
disturbance grows, indicating instability. As usual, the contribution 
coming from the surface tension term has a stabilizing nature ($\sigma$ 
stabilizes modes of large $n$). The factor $(n^{2} - 1)$ in equation~(\ref{dispersion}) arises directly from the 
first order terms in $\zeta$ present in the curvature, while the overall 
factor of $n$ can be traced to the fact that in the generalized 
Darcy's law~(\ref{Darcy}) the velocity is proportional to 
gradients of an effective pressure. With these considerations in mind, 
let us focus on the relation between rotation and 
magnetic field. As a result of centrifugal forcing $N_{\Omega}$ 
may be either positive or negative, depending on the relative 
values of the fluid's densities. If the inner fluid is more 
dense $(\rho_{1} > \rho_{2})$, 
$N_{\Omega} > 0$ and rotation plays a destabilizing role. The opposite 
effect arises when $\rho_{1} < \rho_{2}$. On the other hand, the azimuthal 
magnetic field contribution $N_{B}$ always tends to stabilize the 
interface. This indicates that the rotation-driven instability 
could be delayed or even prevented if a sufficiently strong 
non-uniform, azimuthal magnetic field is applied in the plane 
of the flow. 

A physical explanation for the stabilizing role of the magnetic 
field can be given based on its symmetry properties and 
non-uniform character. Notice that such a field possesses a 
radial gradient. The magnetic field influence is 
manifested as the existence of a 
body force due to field nonuniformity. The field produces 
a force directed radially inward, that tends to move the ferrofluid 
toward the current-carrying wire (regions of higher magnetic field). 
This force opposes the centrifugal force and favors interface 
stabilization. This effect is similar to the gradient-field 
stabilization mechanism discussed by Rosensweig~\cite{Ros} for 
inviscid three dimensional fluid flow problems, and by Zahn and 
Rosensweig~\cite{Za} for viscous, unconfined ferrofluids.

In order to illustrate the role of the magnetic effects in the linear 
stages of the interface evolution, we plot in figure 2 the dimensionless 
growth rate $\bar{\lambda}(n)=[12 (\eta_{1} + \eta_{2})R^{3}/b^{2}\sigma] ~\lambda(n)$ as a function of the mode number $n$, taking 
$N_{\Omega}$=200 and $N_{B}$=0, 100, 200. As expected, for zero applied 
field ($N_{B}=0$) the interface is unstable. If we increase the magnitude 
of the applied field ($N_{B}=100$), even though we get a narrower band 
of unstable modes, the interface remains unstable. 
If we keep increasing the magnetic field intensity, $\bar{\lambda}(n)$ 
becomes negative for all $n > 1$, and the interface 
tends to stabilize. In figure 2 we see that for $N_{B}=200$ 
the interface is stable, due to the action of the azimuthal magnetic field.

A relevant physical quantity can be extracted from the linear 
growth rate: the fastest growing mode $n^{\ast}$, 
given by the closest integer to the maximum of 
equation~(\ref{dispersion}) with respect to $n$ (defined by 
setting $d\lambda(n)/dn=0$), 
\begin{equation}
\label{fastest}
n_{max}=\sqrt{\frac{1}{3} \left [1 + (N_{\Omega} - N_{B}) \right ]}.
\end{equation}
As discussed by Carrillo et al.~\cite{Car} for the non-magnetic, rotating 
case, $n^{\ast}$ is strongly correlated to the number of ripples present 
in the early stages of pattern formation. With the help of their experiments, 
the authors in~\cite{Car} compared the number of ripples with 
$n^{\ast}$ and found a remarkable agreement. Taking this fact into 
account, and inspecting equation~(\ref{fastest}) we verify that, for 
positive $N_{\Omega}$, an increasingly larger $N_{B}$ does not only 
tend to decrease the finger growth rate, but also tends to decrease 
the number of interface ripples. The azimuthal magnetic field (or correspondingly, $N_{B}$) can be seen as a control parameter to 
discipline the number of interface undulations. 

We conclude by contrasting the results obtained above (non-uniform, 
tangential field) with those which arise when a {\it uniform} 
external magnetic field is applied {\it perpendicular} to a 
rotating Hele-Shaw cell containing ferrofluid. By performing 
the linear stability analysis of the system, and using the 
closed form expressions for the 
magnetic term, recently derived in reference~\cite{Eu}, we obtain 
the following linear growth rate
\begin{equation}
\label{perp}
\lambda^{\perp}(n)=\frac{b^{2} \sigma n}{12(\eta_{1} + \eta_{2})R^{3} } \left [ N_{\Omega} + {\cal D}_{n}(p)N_{B}^{\perp} - (n^{2} - 1) \right ],
\end{equation}
where
\begin{equation}
\label{finaldnp}
{\cal D}_n(p)=
{\frac{p^2}{2} \left \{ \left [\psi \left (n + \frac{1}{2} \right ) 
- \psi \left (\frac{3}{2} \right ) \right ] + \left [ Q_{n - 1/2} \left (\frac{p^2 + 2}{p^2} \right ) 
- Q_{1/2} \left (\frac{p^2 + 2}{p^2} \right ) \right ] \right \} },
\end{equation}
$N_{B}^{\perp}=\mu_{0} M^{2}b/2 \pi \sigma$ is the magnetic bond number 
for the perpendicular field configuration, and $p=2R/b$ is 
the aspect ratio. The aspect ratio $p$ should not be confused with 
the pressure. $Q_{n}$ represents the Legendre function of 
the second kind, while the Euler's psi 
function $\psi$ is the logarithmic 
derivative of the Gamma function~\cite{Mag}. Notice that the 
function ${\cal D}_n(p) \ge 0$ for $n > 0$. In contrast to the 
non-uniform, azimuthal applied field discussed earlier, a uniform, 
perpendicular magnetic field tends to destabilize the interface. 
If the interface is already unstable with respect to rotations 
($N_{\Omega}>0$), the introduction of the magnetic field increases 
the interface instability even further. On the other hand, if 
the outer fluid is more dense ($N_{\Omega}<0$), 
the interface can still be deformed by the action of a 
perpendicular magnetic field. 

By comparing the perpendicular field growth rate expression~(\ref{perp}) 
with its azimuthal field counterpart~(\ref{dispersion}), we notice 
that the presence of the function ${\cal D}_n(p)$ in~(\ref{perp}) 
increases the complexity of the problem. Simple expressions 
for the azimuthal case may become not so simple in the perpendicular 
field situation. Consider, for example, the calculation of the fastest 
growing mode $n^{\ast}$. In contrast to the simple expression for 
$n_{max}$ obtained in the azimuthal case (see equation~(\ref{fastest})), 
the equivalent expression for the perpendicular configuration 
cannot be written in a simple form. It is now given by the 
solution of the transcendental 
equation
\begin{equation}
\label{fastest2}
n^{2}=\frac{1}{3} \left [ 1 + N_{\Omega} + \frac{\partial }{\partial n} 
\Big ( n~{\cal D}_n(p) \Big ) ~N_{B}^{\perp} \right ].
\end{equation}
Numerical evaluation of equation~(\ref{fastest2}) shows that 
$n_{max}$ (and consequently, $n^{\ast}$) increases with 
the magnitude of the perpendicular applied 
field. In this case, larger values of $N_{B}^{\perp}$ tend to increase 
the number of interface ripples (see figure 3).  

In summary, we have shown that the inclusion of magnetic effects 
into the rotating cell problem provides new mechanisms for 
stabilizing/destabilizing the interface. Based on the 
relative simplicity of the experimental Hele-Shaw setup, 
it would be of considerable interest to perform experiments in 
Hele-Shaw cells simultaneously including centrifugal, magnetic and 
injection driving. The combination of these effects is likely to lead 
to new and exciting interfacial patterns in the highly nonlinear regime.

This work was supported by CNPq and FINEP.

\pagebreak
\noindent
\centerline{{\large {FIGURE CAPTIONS}}}
\vskip 0.5 in
\noindent
{FIG. 1:} Schematic configuration of the rotational Hele-Shaw flow 
with ferrofluid.

\vskip 0.25 in
\noindent
{FIG. 2:} Variation of the dimensionless growth rate 
$\bar{\lambda}(n)=[12 (\eta_{1} + \eta_{2})R^{3}/b^{2}\sigma] ~\lambda(n)$ 
as a function of $n$  for $N_{\Omega}=200$ and (a) $N_{B}=0$, 
(b) $N_{B}=100$, and (c) $N_{B}=200$. The peak location and width of the 
band of unstable modes decrease with increasing $N_{B}$.

\vskip 0.25 in
\noindent
{FIG. 3:} Plot of $n_{max}$ as a function of $N_{B}^{\perp}$ for increasing 
values of $N_{\Omega}$, and $p=20$. These curves are obtained by 
numerically solving equation~(\ref{fastest2}).

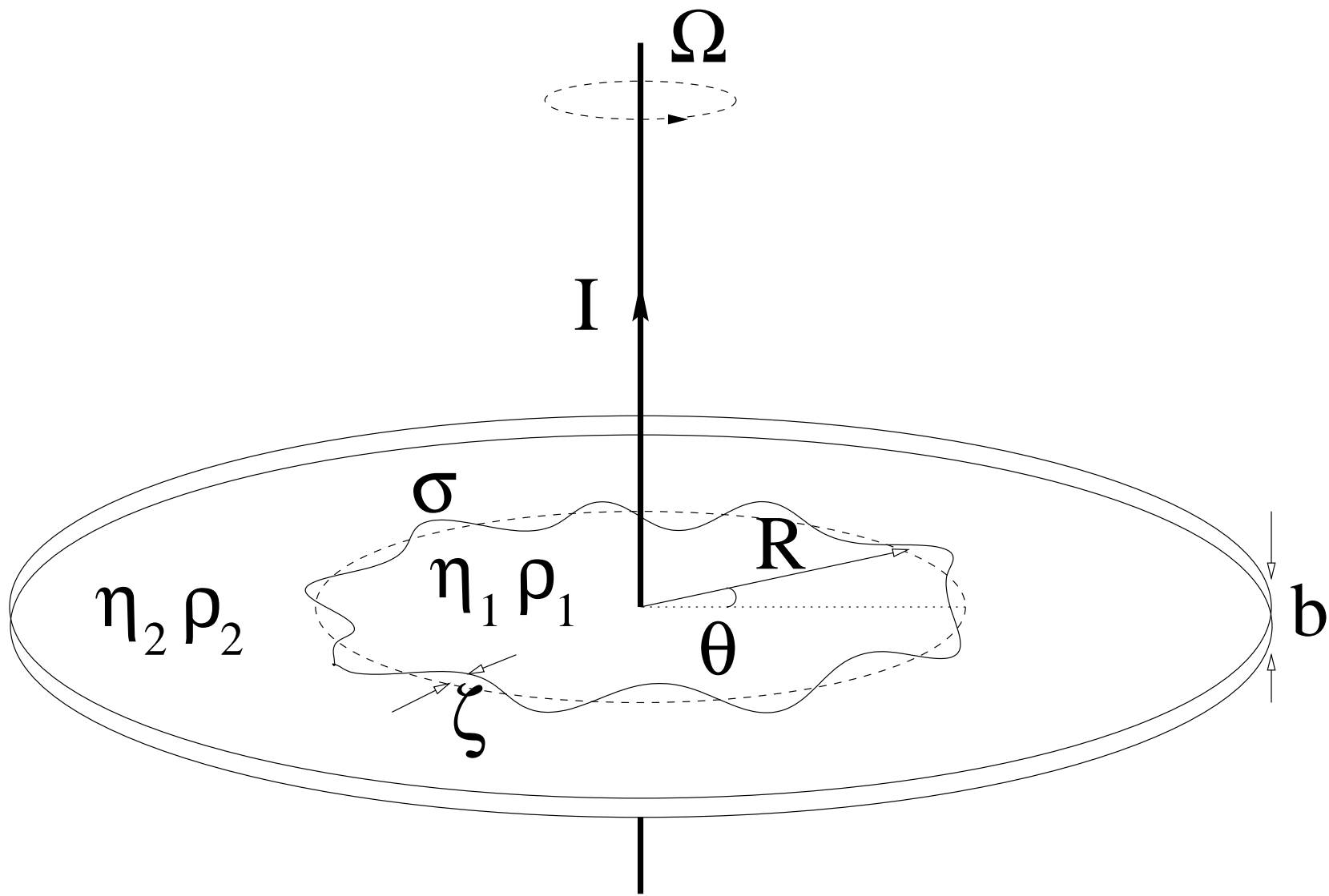
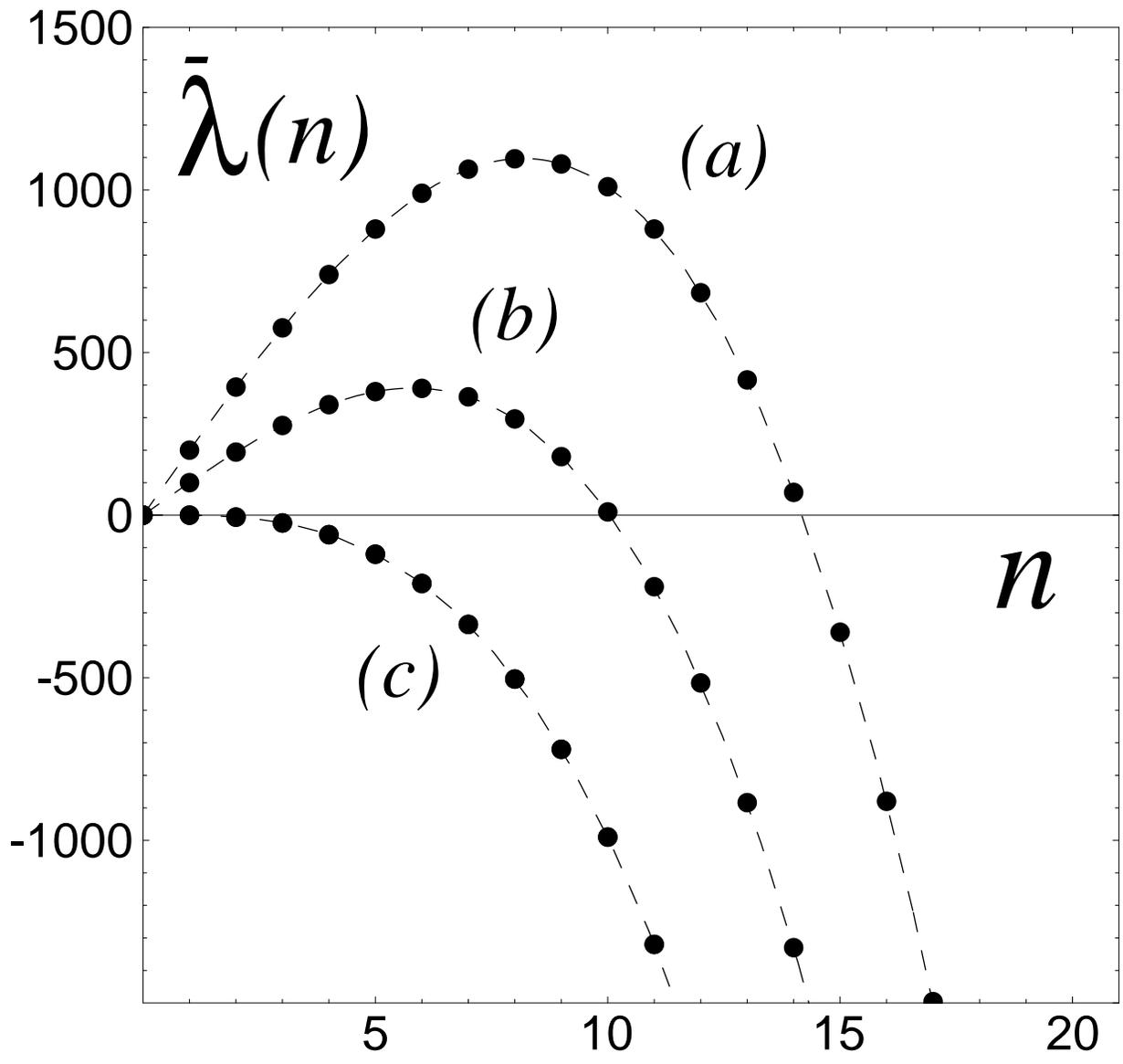
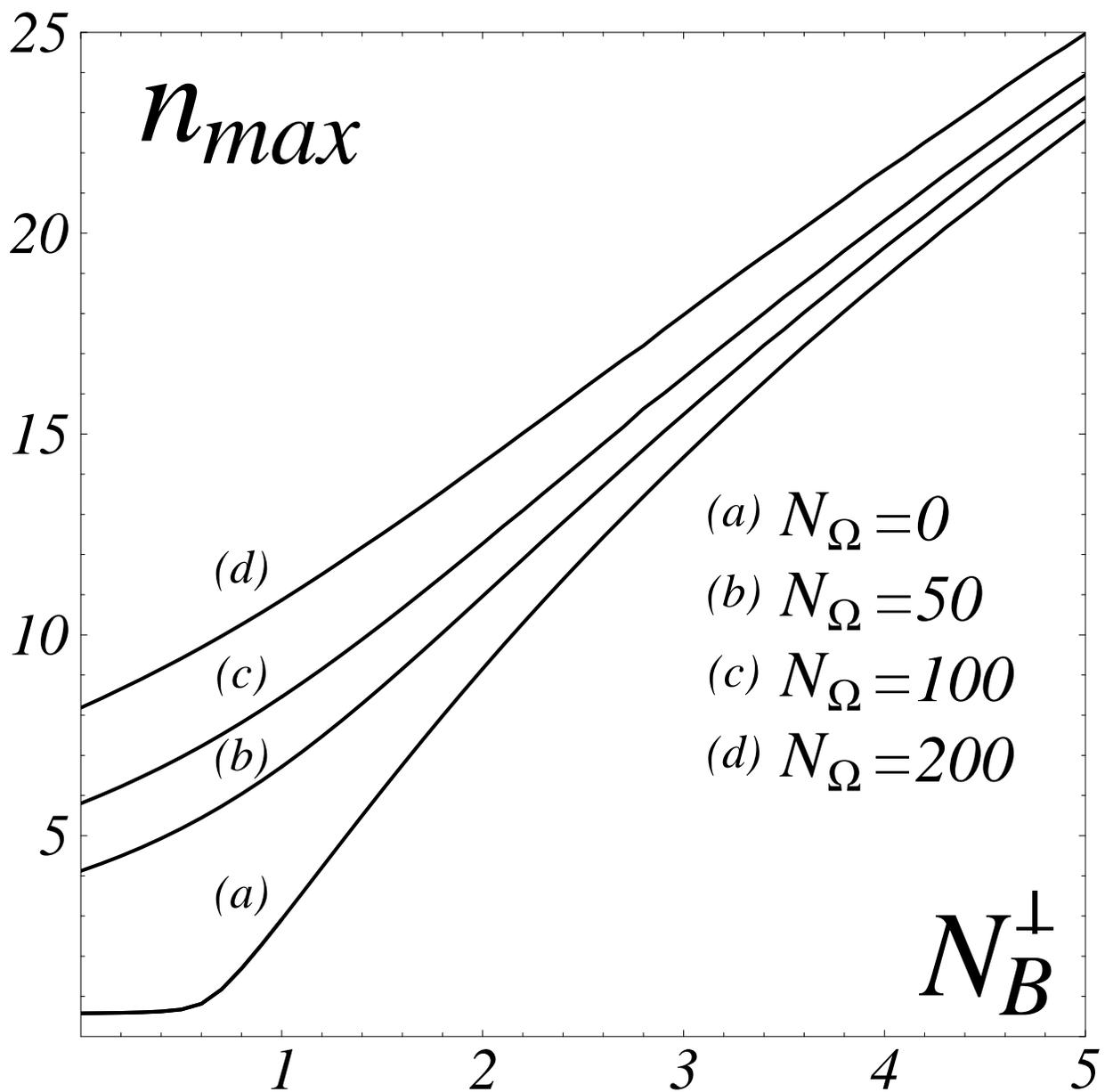
\end{document}